\documentclass[prl,twocolumn,showpacs,amsmath,amssymb,floatfix]{revtex4}
\usepackage{graphicx}
\usepackage{dcolumn}

\begin{document}

\title{Relaxation dynamics of an elastic string in random media}

\author{Jae Dong Noh }
\affiliation{Department of Physics, University of Seoul,
  Seoul 130-743, Korea}
\author{Hyunggyu Park }
\affiliation{School of Physics, Korea Institute for Advanced Study, Seoul
130-722, Korea}
\date{\today}

\begin{abstract}
We investigate numerically the relaxation dynamics
of an elastic string in two-dimensional random media by thermal fluctuations
starting from a flat configuration. Measuring spatial fluctuations of its mean position,
we find that the correlation length grows in time asymptotically as $\xi \sim (\ln t)^{1/\tilde\chi}$.
This implies that the relaxation dynamics is driven by thermal activations over random energy
barriers which scale as $E_B(\ell) \sim \ell^{\tilde\chi}$ with
a length scale $\ell$. Numerical data strongly suggest that the energy barrier
exponent $\tilde{\chi}$ is identical to the energy fluctuation exponent
$\chi=1/3$. We also find that there exists a long transient regime, where the correlation length follows a power-law dynamics as
$\xi \sim t^{1/z}$ with a nonuniversal dynamic exponent $z$.
The origin of the transient scaling behavior is discussed in the context of
the relaxation dynamics on finite ramified clusters of disorder.

\end{abstract}
\pacs{05.70.Ln, 64.60.Ht, 75.50.Lk, 75.60.Ch}

\maketitle

Interaction and quenched disorder are essential ingredients in condensed
matter physics. Many-body systems may undergo a phase transition as a
cooperative phenomenon mediated by interaction. When quenched disorder comes
into play, there may emerge a glass phase in which degrees of
freedom are pinned by random impurities and
a slow relaxation dynamics appears.
An elastic string in random media
is one of the simplest systems
where the interplay between interaction and quenched disorder yields a
nontrivial effect~\cite{Kardar87,Fisher91}.
This has been studied extensively in literatures since it
is relevant to many interesting physical systems such as a growing
interface~\cite{KPZ86},
a domain wall in random magnets~\cite{Huse85,Bray9402}, and
a magnetic flux line in superconductors~\cite{Kardar97}.

In low temperatures, one may approximate an elastic string as an
elastically coupled directed polymer where no overhang is allowed.
Then it can be described by a single valued function $\bm{x}(u)$,
where $\bm{x} \in \mathcal{R}^d$ and $u \in \mathcal{R}$
are the transverse and the longitudinal coordinates to the polymer direction, respectively, in
a $(d+1)$ dimensional space. The energy of a polymer of length $L$
in a configuration $\bm{x}(u)$ with $0\leq u \leq L$
is given by the Hamiltonian
\begin{equation}
\mathcal{H} = \int_0^L du \left[ \frac{1}{2}
\left| \frac{\partial \bm{x}}{\partial u}\right|^2 +
V(u,\bm{x}(u))\right] \ .
\end{equation}
The first term accounts for an elastic tension and the second term
$V(u,\bm{x})$ is a random pinning potential with short-range
correlations.

Equilibrium properties of the directed polymer in random media (DPRM) are rather well understood.
The tension favors a flat state, while thermal fluctuations
and the disorder potential favor a rough state.
The competition between them  leads to the scaling law
$|\Delta {\bf x}| \sim L^{\zeta}$ for the
transverse fluctuation (interface roughness) and $\Delta E \sim L^{\chi}$
for the (free) energy fluctuation.
The quenched disorder is relevant for $d\leq 2$, and the polymer is in a
super-rough phase ($\zeta > 1/2$) at all temperatures. Especially for $d=1$, the scaling exponents
are known exactly as $\zeta_{1D} = 2/3$ and $\chi_{1D} = 1/3$~\cite{Huse85}.
For $d>2$, it is believed that there is a transition from a super-rough phase
into a normal-rough phase ($\zeta = 1/2$) as the temperature $T$ increases~\cite{Fisher91}.
In the latter, the thermal fluctuations dominate over the disorder fluctuations,
while vice versa in the former.

When a polymer is in a nonequilibrium state, e.g., a flat configuration, it
will relax to the equilibrium rough state.
Without disorder or in the normal-rough phase with disorder for $d>2$, the elastic polymer equilibrates diffusively.
The correlation length $\xi$ in the longitudinal $u$ direction grows
algebraically in time as $\xi \sim t^{1/z_o}$ with the dynamic exponent $z_o=2$.
In the presence of the quenched disorder, one expects a slower relaxation
because random impurities tend to trap the polymer into metastable states in local energy valleys.
Upon equilibration, the polymer has to overcome energy barriers $E_B$ separating those valleys through
thermal fluctuations to approach the true equilibrium state.
It is believed that the energy barrier height scales  as $E_B(\ell) \sim \ell^{\tilde\chi}$
in a region with linear size $\ell$.

Thermal activations allow the correlated polymer segment of length $\xi$ to overcome the energy barriers in a time scale $t_\xi \sim e^{E_B(\xi)/T}$. Then, it follows that the correlation length $\xi$ grows as
\begin{equation}\label{eq:corr_length}
\xi(t) \sim (T\ln t)^{1/\tilde\chi}, \
\end{equation}
with the universal energy barrier
exponent $\tilde\chi$, independent of the disorder strength and the temperature~\cite{Huse85}.
Assuming that there is only a single relevant energy scale in this system,
the exponent $\tilde\chi$ should be equal to the energy
fluctuation exponent $\chi$~\cite{Huse85}. This conjecture is supported at least in low dimensions~\cite{Drossel95}.

However, even for
$d=1$, there is a long-standing controversy on the scaling law of Eq.~(\ref{eq:corr_length}).
Numerical simulation study~\cite{Kolton05}
reports a signature of the expected logarithmic scaling, but only after a long and clean intermediate
power-law scaling regime where $\xi\sim t^{1/z}$
with a nonuniversal dynamic exponent $z$, whose origin is not clear. Moreover, the scaling exponent associated
with the logarithmic scaling seems different from the conjectured value of $\tilde\chi=\chi=1/3$~\cite{Kolton05}.
There is also a recent claim of $\tilde\chi=d/2$ based on the droplet theory~\cite{Monthus08}.
Besides, there are many numerical works in the context of domain wall coarsening dynamics
in two dimensional random ferromagnets~\cite{Bray9402}, which seem to support the nonuniversal power-law scaling without any signature of the asymptotic logarithmic
scaling~\cite{OC86,PPR04}. These intriguing results are also left unexplained.

In this work, we study the relaxation dynamics of the DPRM in $(1+1)$ dimensional lattices
with extensive numerical simulations and a scaling theory approach. Measured are the spatial fluctuations of the mean position of the polymer, from which we
derive the dynamic scaling behavior of the correlation length $\xi(t)$ through a simple scaling hypothesis.
The purpose of this study is to settle down the controversy by providing a decisive numerical evidence on the asymptotic relaxation dynamics of the DPRM for $d=1$. In addition, we suggest a reasonable scenario for the origin of the transient power-law scaling regime.

We consider a discrete model for the DPRM in the 45-degree-rotated square
lattices of size $L\times M$. Each lattice site is
represented as $(i,x)$ with the longitudinal coordinate $i=0,\cdots,L-1$ and
the transverse coordinate $x=-M/2+1,\cdots,M/2$ with the constraint $i=x$ in modulo 2.
Assigned to bonds are quenched disorder variables $J$ which are distributed independently and randomly
according to a probability density function $p(J)$.
The polymer of length $L$ is placed along the bonds and directed in the longitudinal direction without
any back bending. Then its configuration is described by the fluctuating variables
$\{x(i)\}$ with the solid-on-solid (SOS) constraint of $|x(i)-x(i\pm 1)|=1$ for all $i$.
We adopt the periodic boundary condition in the longitudinal direction,
i.e., $x(L) = x(0)$. The transverse size $M$ is taken to be large
enough ($M=4096 \sim 8192$) to avoid any possible interference.

The polymer energy is given by the lattice Hamiltonian
\begin{equation}\label{Hamiltonian}
\mathcal{H} = \sum_{i=1}^L J(i,x(i);i+1,x(i+1)) \ ,
\end{equation}
where $J(i,x;i+1,x')$ denotes the disorder strength of the bond between neighboring
sites $(i,x)$ and $(i+1,x')$.
In this study, we consider the uniform distribution in the range
$-1\leq J\leq 1$ and the bimodal distribution
$p(J) = f \delta(J+1/2) + (1-f) \delta (J-1/2)$ with a model parameter $f$.
It turns out that both cases lead to the same conclusion.

We start with the flat configuration with $\{x(i) = i\mod{2}\}$ at $t=0$ and study its
relaxation dynamics toward the equilibrium state
at temperature $T$. We adopt the Glauber Monte Carlo dynamics:
First select a site $i$ at random and try to flip $x(i)
\rightarrow x(i)\pm 2$ with probability $1/2$, respectively. Unless the
trial violates the SOS constraint, it is accepted with the probability of
$\min[1,e^{-\Delta E/T}]$ where $\Delta E$ is the energy change. The time is
incremented by one unit after $L$ such trials.

We focus on the spatial dispersion of the mean position
$\overline{x} \equiv \sum_{i} x(i)/L$ as function of elapsed time $t$, which is given as
\begin{equation}\label{eq:disp_x}
(\Delta x)^2 (t) \equiv \left[ \left\langle \left(\overline{x}(t) -
\overline{x}(0)\right)^2  \right\rangle_T\right]_D \ .
\end{equation}
Here $\langle \cdot\rangle_T$ and $[ \cdot ]_D$ denote the thermal and
disorder average, respectively.
Let $\xi(t)$ be the characteristic correlation length of the polymer
in the longitudinal direction.  At time $t$, each segment of size $\xi(t)$
equilibrates with transverse displacement of the order
$\delta x \sim \xi^\zeta$ with the roughness exponent $\zeta$.
When $\xi \ll L$, each segment is independent and
the total displacement is given by
\begin{equation}\label{eq:scaling_form}
(\Delta x)^2 \sim \frac{ \xi^{2\zeta} }{ (L/\xi) }
\sim \frac{\xi^{1+2\zeta}}{L} \ .
\end{equation}
Utilizing this relation, we can derive the correlation length $\xi(t)$ from
the ensemble-averaged global quantity $(\Delta x)^2(t)$, which usually bears a better
statistics than the distance-dependent correlation function of the
transverse displacement.

When the polymer fully equilibrates, i.e., $\xi(t=\tau) \simeq L$ with the
relaxation time $\tau=\tau(L)$, the polymer as a whole (the mean position) starts to diffuse normally in the transverse direction.
One expects that $(\Delta x)^2$ grows linearly in time scaled by $\tau$ as
\begin{equation}\label{eq:scaling_form_s}
(\Delta x)^2 \sim L^{2\zeta} \left(\frac{t}{\tau}\right) \ .
\end{equation}
Without disorder, the motion of the polymer is governed by the linear
Edward-Wilkinson~(EW) equation~\cite{EW82}.
The EW class is characterized by $\zeta=1/2$
and  $\xi \sim t^{1/2}$, i.e., $\tau\sim L^2$.
In this case, the dynamic behaviors before and after equilibration,
Eqs.~(\ref{eq:scaling_form}) and (\ref{eq:scaling_form_s}), follow the same
scaling law $(\Delta x)^2 \sim t/L$ at all $t$.
Indeed, this coincides with the exact solution of the EW equation. However, in general with disorder,
these two scaling laws are distinct.

\begin{figure}[t]
\includegraphics*[width=\columnwidth]{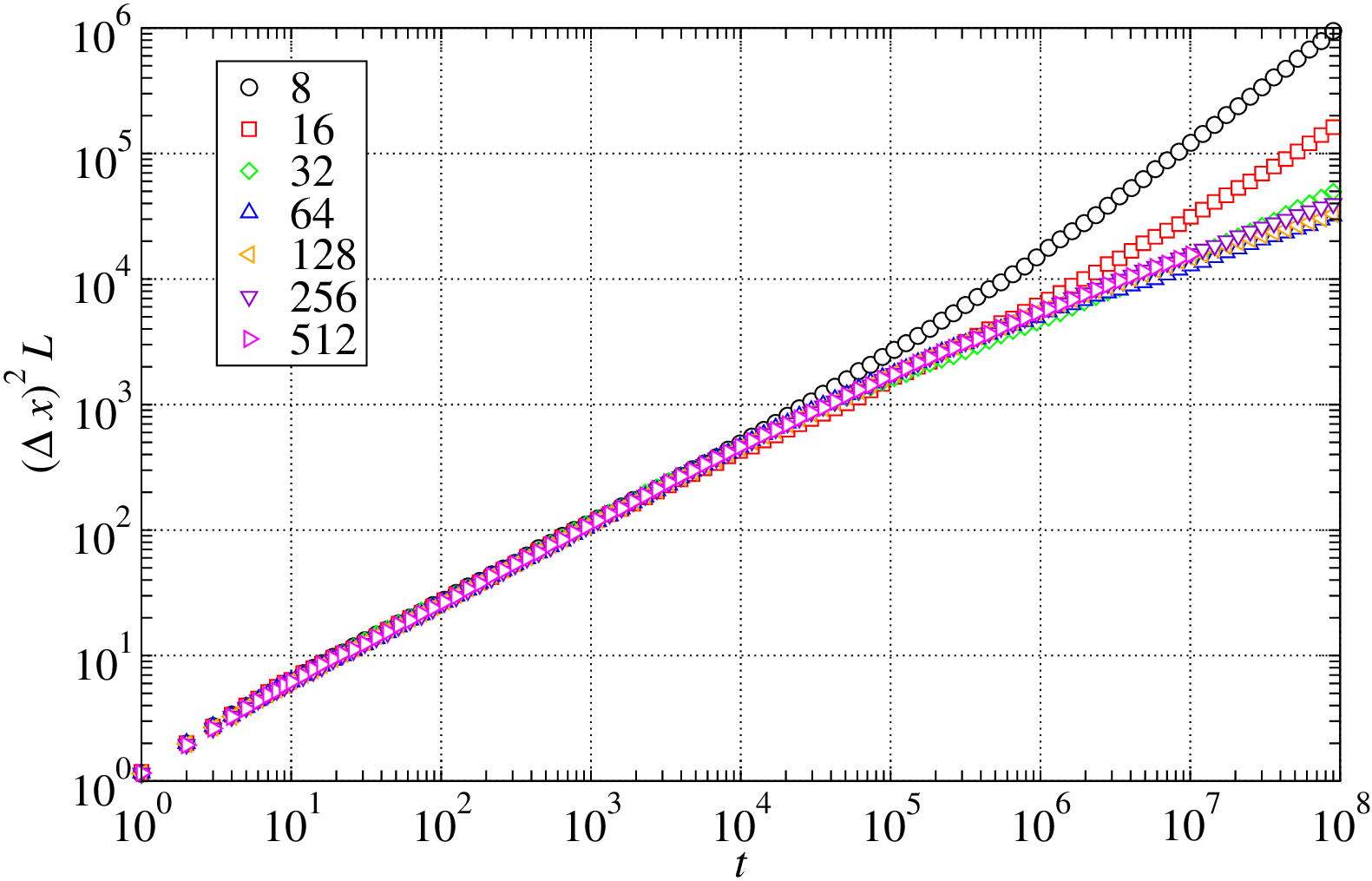}
\caption{Monte Carlo simulation data for the system with the bimodal
disorder distribution with $f=0.25$ and at $T=1.0$.
Different symbols represent data for different polymer lengths $L$.
The data are averaged
over $N_S=5000$ disorder samples.}\label{fig1}
\end{figure}

\begin{figure}[b]
\includegraphics*[width=\columnwidth]{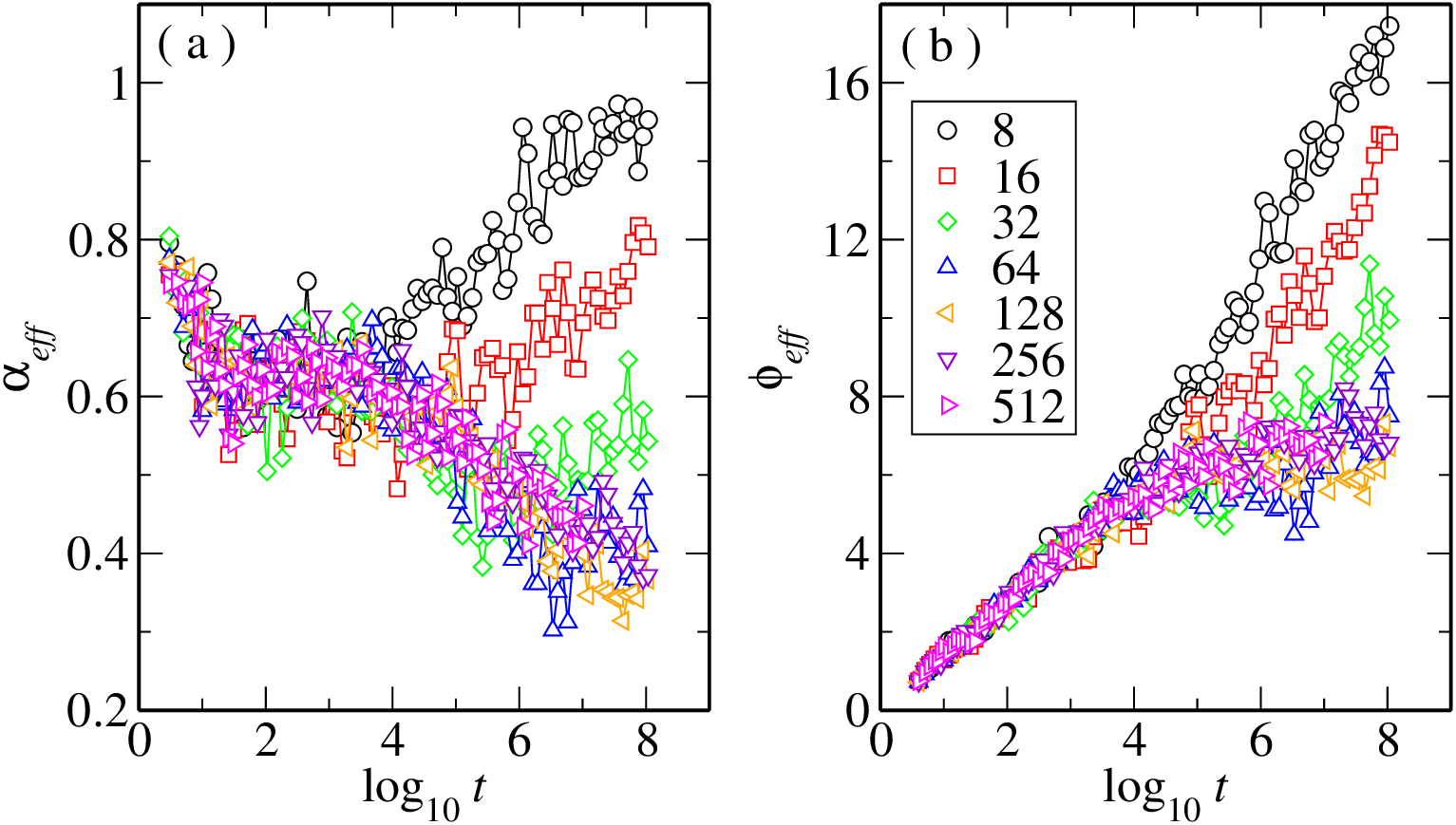}
\caption{Effective exponent plots for the data given in Fig.~\ref{fig1}.
Our estimates are $\alpha=0.63\pm 0.03$ in regime II and
$\phi=6.8\pm 0.5$ in regime III.}
\label{fig2}
\end{figure}

We have performed extensive Monte Carlo simulations to examine the scaling
property of $(\Delta x)^2$. Figure \ref{fig1} shows a plot of the numerical data with
the bimodal disorder distribution with $f=0.25$ at $T=1.0$.
As the scaling form predicts in Eq.~(\ref{eq:scaling_form}), $(\Delta x)^2$ is
inversely proportional to $L$ for $t<\tau(L)$, so it is convenient to plot
$L (\Delta x)^2$ versus $t$, where all curves with different $L$ collapse into one
scaling  curve for $t<\tau(L)$ and then start to deviate and show
the finite size effects given by Eq.~(\ref{eq:scaling_form_s}).

We find that there exist four distinct regimes:
(I) For $t<t_0 (\sim 10^1)$, the polymer moves
diffusively as $L (\Delta x)^2 \sim t$. In this regime, the polymer
behaves as in the EW class since it does not feel the disorder pinning as yet.
(II) For $t_0 < t < t_c(\sim 10^{4 \sim 5}) $,
the polymer is affected by the disorder and exhibits a power-law scaling behavior
as $L (\Delta x)^2 \sim t^{\alpha}$ with a nonuniversal exponent $\alpha$.
The crossover time $t_c$ is very large but finite and independent of $L$, which implies that
this power-law scaling is transient.
(III) For $t_c < t < \tau(L)$, there is a continuous downward curvature in the plot
suggesting a possible logarithmic scaling with $L (\Delta x)^2
\sim (\ln t)^\phi$.
The crossover time $\tau(L)$ increases indefinitely with $L$, which implies that
this regime should be the true asymptotic scaling regime.
(IV) For $t> \tau(L)$, the polymer displays a diffusive motion with
a size-dependent diffusion amplitude as in Eq.~(\ref{eq:scaling_form_s}).

We investigate the scaling behavior in each regime quantitatively. Useful
are the effective exponents defined as
$\alpha_{eff}(t) \equiv d \ln (L (\Delta x)^2 ) / d \ln t$ and
$\phi_{eff}(t) \equiv d
\ln (L (\Delta x)^2 ) / d \ln \ln t$. If $L (\Delta x)^2 \sim t^{\alpha}$ as in the regime II,
one would obtain that $\alpha_{eff}(t) =
\alpha$ and $\phi_{eff}(t) = \alpha \ln t$. On the other hand, if
$L (\Delta x)^2 \sim (\ln t)^\phi$ as in the regime III,
one would obtain that $\alpha_{eff}(t) = \phi / \ln t$ and $\phi_{eff}(t) =
\phi$.

We plot the effective exponents in Figs.~\ref{fig2} (a) and (b).
In (a), one can clearly see a plateau at $\alpha_{eff} = 0.63\pm 0.03$ for $10^1 \lesssim
t \lesssim 10^4(=t_c)$ (regime II). For $t>t_c$, it continuously decreases
in the regime III before hiking up in the regime IV as expected. In (b),
there appears a plateau
at $\phi_{eff} = 6.8 \pm 0.5$ in the regime III, which widens as $L$ increases.
These numerical evidences lead to a definitive conclusion that the asymptotic motion of the polymer
follows the logarithmic (not power-law) scaling as
\begin{equation}\label{eq:log_scaling_Dx}
L (\Delta x)^2 (t) \sim (\ln t)^\phi \ ,
\end{equation}
with the exponent $\phi=6.8\pm 0.5$. In terms of the correlation length,
we find, using  Eq.~(\ref{eq:scaling_form}),
\begin{equation}\label{eq:log_scaling_xi}
\xi (t) \sim (\ln t)^{1/\tilde\chi} \ ,
\end{equation}
where $1/\tilde\chi=\phi/(1+2 \zeta)=2.9\pm0.2$. The estimated value of
$\tilde\chi=0.34\pm 0.03$ clearly favors the conjecture value of $\tilde\chi=\chi=1/3$
and invalidates the recent claim of $\tilde\chi=1/2$ (or equivalently $\phi=14/3$).

\begin{figure}[t]
\includegraphics*[width=\columnwidth]{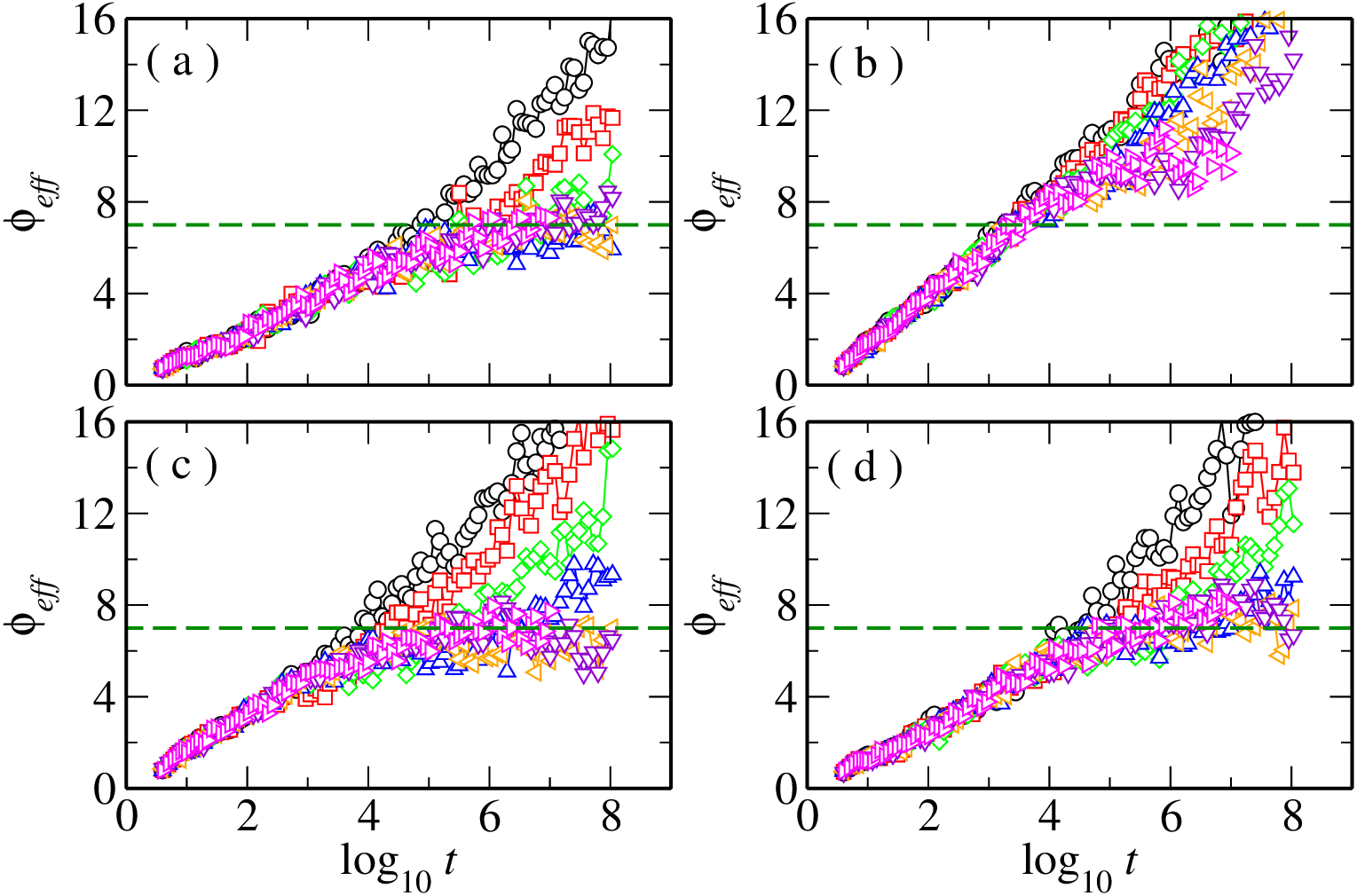}
\caption{Effective exponent plots with the uniform disorder distributions
with the temperature $T=0.25$ in (a) and $T=0.75$ in (b),
and the bimodal disorder distribution with $f=0.1$ and $T=1$ in (c)
and $f=0.5$ and $T=1$ in (d). The symbols have the same attribute as in
Fig.~\ref{fig1}. The dashed lines are drawn at
$\phi_{eff}=7$.}
\label{fig3}
\end{figure}

In order to examine the universality of the logarithmic scaling behavior,
we have performed the simulations with different values of $f$
in the case of the bimodal disorder distribution and also in the case of the uniform disorder
distribution. The effective exponents are then presented in
Fig.~\ref{fig3}. The plateaus around $\phi_{eff}=7.0$ shown in
Figs.~\ref{fig3}(a), (c), and (d) confirm the universality of the logarithmic scaling
as well as the universality of its exponent $\tilde\chi$.
The plot in Fig.~\ref{fig3}(b) seems to be incompatible with $\phi=7$.
Comparing Figs.~\ref{fig3}(a) and (b),
one can notice that the finite-size behavior (regime IV) sets in earlier at the higher
temperature. So the logarithmic scaling regime is observed only when $L\ge 256$ in (b),
while it is already evident at $L= 64$ in the other cases.
This suggests that one would need larger polymers at higher temperatures.

It is puzzling why there exists the extremely long transient regime II where
the polymer relaxation seems to follow a power-law scaling such as
$L (\Delta x)^2 \sim t^\alpha$ or $\xi \sim t^{1/z}$ with $z=(1+2\zeta)/\alpha$. Moreover
the exponent $\alpha$ is nonuniversal and varies with the disorder strength and the
temperature. Such a transient behavior was also reported in Ref.~\cite{Kolton05,power-law-literatures},
but its origin has never been explored.
We suggest one reasonable scenario as below.

It is convenient to consider the bimodal disorder distribution.
When $f< 1/2$, the energetically favorable bonds with $J=-1$ may play
the role of local pinning centers for the polymer~\cite{comment2}.
As the polymer considered here is directed, relevant are the directed percolation clusters
of the pinning bonds. These clusters are ramified but finite in size, as
the bond density $f$ is smaller than the directed percolation threshold
$f_c \simeq 0.6449$ in the square lattice~\cite{Jensen99}.
The characteristic size and the mean distance between them are denoted
by $l_0$ and $l_1$, respectively. After the initial diffusive motion,
polymer segments are trapped by those clusters independently as long
as the correlation length is smaller than the cluster size~($\xi <l_0$).

The pinning mechanism
in fractal-like ramified lattices is different from that in bulks.
The energy barrier height in such lattices is shown to scale logarithmically
with a length scale $\ell$ as $E_B(\ell) \simeq E_0 \ln \ell$ with an universal constant
$E_0$ depending only on the ramification degree~\cite{Henley85}.
Then, the time scale associated with the
thermal activation of the correlated segment of length $\xi$
is given by
$t_\xi \sim e^{E_B(\xi)/T}\sim \xi^{E_0/T}$. This yields the power-law
growth of the correlation length as $\xi \sim \sqrt{t/t_\xi} \sim
t^{1/z}$ with the nonuniversal dynamic exponent $z=2+E_0/T$. The temperature
dependence seems consistent with our numerical estimates for $z$ (not shown here).
When $\xi$ exceeds $l_0$, the polymer segments are pinned by a few pinning
clusters. If $\xi\ge l_1$, then the polymer starts to be pinned
collectively, and the transient power-law scaling behavior crosses over to
the asymptotic logarithmic scaling behavior.

Finally, we add one remark on the domain coarsening dynamics in the two-dimensional
random ferromagnets. When the system is quenched well below
an ordering temperature from a disordered state,
the characteristic size $R$ of ordered domains increases and the domain wall motion may be described by the DPRM.
Hence, it is natural to expect that $R(t) \sim \xi^{2-\zeta}\sim
(\ln t)^{(2-\zeta)/\tilde\chi}$~\cite{Huse85}.
Surprisingly, recent high accuracy numerical simulation studies report that
$R(t) \sim t^{1/z}$ with a nonuniversal exponent $z$~\cite{PPR04}.
Our result suggests that those behaviors
may be due to the pinning of domain walls by finite pinning clusters
in the transient regime.

In summary, we have investigated numerically the relaxation dynamics of the
DPRM.
The numerical data show unambiguously that the correlation length grows as
$\xi \sim t^{1/z}$ in the transient regime and then
$\xi\sim (\ln t)^{1/\tilde\chi}$ in the asymptotic regime.
The transient behavior is originated from the pinning independently by local ramified impurity
clusters. The asymptotic logarithmic scaling is compatible with the scaling
picture that the energy barrier height scales in the same way as the energy fluctuations
with $\tilde\chi=\chi=1/3$.
Implication on the domain coarsening dynamics is also discussed.

We thank Doochul Kim, Malte Henkel, and Heiko Rieger for useful discussions.
This work was supported by KOSEF grant
Acceleration Research (CNRC) (Grant No. R17-2007-073-01001-0).


\begin{references}
\bibitem{comment2} When $f>1/2$,  the energetically favorable bonds become
the majority and the unfavorable bonds play the role of local pinning centers.
Although the correspondence is not exact, we expect that the whole arguments developed for small $f$
are also valid for large $f$.
\end{references}
\end{document}